# Dose Selection Balancing Efficacy and Toxicity using Bayesian Model Averaging

A. Lawrence Gould
Merck & Co., Inc., Rahway, NJ, USA
24 January 2024

**ABSTRACT**

Successful pharmaceutical drug development requires finding doses that provide an 'optimum' balance between efficacy and toxicity. Competing responses to 'dose' such as efficacy and toxicity often will increase with 'dose', and it is important to identify a range of doses to provide an acceptable efficacy response (minimum effective dose) while not causing unacceptable intolerance or toxicity (maximum tolerated dose). The BMA-Mod Bayesian model averaging framework provides a general, statistically valid, distributionally flexible, and operationally practical graphically oriented model-agnostic strategy for predicting efficacy and toxicity outcomes both in terms of expected responses and in terms of predictions for individual patients when the functional forms of the efficacy and toxicity dose response relationships are unknown. The performance of the strategy is evaluated via simulation when efficacy and toxicity outcomes are considered marginally, when they are associated via gaussian and Archimedean copulas, and when they are expressed in terms of clinically meaningful categories. In all cases, the BMA-Mod strategy identified consistent ranges of acceptable doses.

## 1 INTRODUCTION

### 1.1 Background

Clinical development programs need to identify intervention levels (typically doses, but possibly treatment duration or dosing frequency or regimen) to evaluate at each developmental stage (intervention is taken as 'dose' in what follows for convenience). Efficacy and toxicity findings from phase 2 dose response trials typically are used to determine the doses to carry forward to large-scale phase 3 programs to avoid expensive late-stage failures because of inadequate efficacy or excess toxicity[1, 2]. Highly effective but ill-tolerated doses are unlikely to be therapeutically useful or commercially viable. Efficacy and toxicity often will increase with 'dose' and it is important to identify a range of doses to provide an acceptable efficacy response (minimum effective dose) while not causing unacceptable intolerance or toxicity (maximum tolerated dose). Figure 1 illustrates this goal. If $AR_E$ and $AR_S$ represent the ranges of doses providing acceptable efficacy and safety responses, respectively, then the range of doses to carry forward in development is their intersection, $AR_{ES} = AR_E \cap AR_S$. Clinical, regulatory, and commercial considerations usually determine which doses in the range are carried forward. Efficacy and toxicity may not be the only criteria defining an acceptable dose range. For example, bioavailability as expressed in the pK/pD profile also may be relevant. In this case, if $AR_B$ represents the range of doses providing acceptable bioavailability, then the range of doses to carry forward is $AR_{BES} = AR_B \cap AR_E \cap AR_S$.

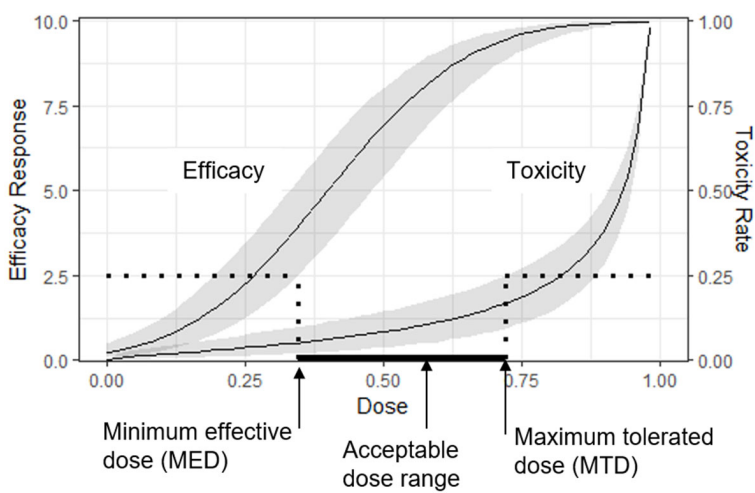

Figure 1. Range of acceptable doses balancing efficacy and toxicity. The shaded regions represent 95% simultaneous credible intervals for the dose-response relationship. The expected efficacy response exceeds 2.5 with posterior probability at least 0.975 for doses exceeding the MED. The expected toxicity rate is less than 25% with posterior probability at least 0.975 for doses less than the MTD.

In reality dose-response modeling strategies may or may not assume specific functional forms for the relationships corresponding to each response. The responses may be continuous, binary, categorical, or counts, with an appropriate link function to relate the response to some function of the 'dose'. Efficacy and toxicity responses often will be associated or correlated, if only because the the active component of any product often affects receptors on organ sites other than the intended therapeutic target.

### 1.2 Modeling Efficacy and Toxicity

De Leon and Wu [3] provide a succinct description of the issue and discussion of the consequences of considering and ignoring the association between efficacy and safety responses at specific dose levels. Their approach uses marginal models to express dose-response relationships and uses a copula to incorporate potential association between efficacy and toxicity outcomes. For the approach considered here, marginal models will suffice to identify doses providing at least adequate efficacy and at least tolerable toxicity. Characterizing efficacy responses as a function of dose conditionally on what toxicity might occur is unlikely to be useful in a real clinical setting because whether toxicity does or does not occur will not be known at the time therapy starts. For perspective, Appendix 3 presents consequences of conditioning efficacy dose-response relationships on whether toxicity events occurred or not.

## 2 PREVIOUS WORK

Methods have been described for designing and analyzing trials that address efficacy and toxicity together. The analysis models include generalized linear mixed models[4], nonparametric models[5], gaussian process models [6], and copula models that incorporate outcomes with known response distributions. [7-9] Nonparametric functional forms also have been suggested for characterizing the dose-response relationship in the marginal distributions of copula models,[10] possibly requiring evaluating dose responses at more doses than conventional trials employ.

### 2.1 Design

There is a substantial literature on the design of trials for determining optimal dosages based on efficacy and toxicity outcomes, how many and which doses are included, sample sizes, etc. Denman *et al* [11] showed that including the dependence between bivariate responses in the design process by using copulas led to more efficient parameter estimates. Dragalin and Fedorov [12, 13] proposed locally optimal adaptive procedures for dose-finding based on efficacy and toxicity using Gumbel logistic regression or a Cox bivariate binary model.[14] Kao and Khogeer [15] used Loewner ordering and Chebycheff systems to identify locally optimal designs for mixed (continuous and binary) outcomes. Ivanova *et al* [16] used a utility function-based adaptive dose finding strategy to select the dose within a defined set with the best efficacy/tolerability profile. Kim and Kao [17] described the use of mathematical programming to get bounds on the number of distinct support points need to obtain locally optimal designs. Schorning *et al* [18] observed that minimally supported approximately locally optimal designs (with a minimum number of dose levels) are efficient if continuous bivariate efficacy and toxicity outcomes are not highly correlated, but do not recommend their use if they are strongly correlated. Takahashi [6] proposed a design based on a Bayesian optimization framework for identifying optimal doses for biologic agents in phase I/II trials by modeling dose-response relationships via nonparametric models utilizing a Gaussian process prior, and including the uncertainty of estimates. Takeda *et al* [19, 20] proposed a generalized model-assisted Bayesian optimal interval design for dose-finding using isotonic regression to model the toxicity outcomes and fractional polynomials for efficacy. Yeung *et al* [5] described dose-escalation designs that incorporate both the dose-limiting events and dose-limiting toxicities (DLTs).

## 2.2 Analysis

Current analysis methods generally assume known dose-response functional forms. A. Tao *et al* [4] extended the MCP-Mod approach [21] to incorporate both efficacy and toxicity information in determining appropriate doses to carry to phase 3. Costa and Drury [7] described the application of a generalized linear model and a bivariate or trivariate Gaussian copula model for jointly modeling benefit and risk in drug development when the benefit was expressed as a normally distributed continuous variable and safety was expressed as one or two variables with Bernoulli distributions. Y. Tao *et al* [9] extended the continuous reassessment method (CRM) to accommodate both efficacy and toxicity outcomes to identify optimal dose selection for phase 3 trials using an Archimedean copula algorithm. Deldossi *et al* [22] compared via simulation Clayton and Gumbel copulas with logistic marginals for binary efficacy and toxicity outcome, and found that a criterion based on the probability of success without toxicity could distinguish between the copulas. Papanikos *et al* [8] compared the result of applying a bivariate random effects meta-analysis and a bivariate Gaussian copula with binomial margins to data on interim and final counts of endpoint events among patients with chronic myeloid leukemia via a Bayesian analysis implemented in the `cmdstan` R package. [23] They concluded that the copula model was the most appropriate approach for assessing study level association patterns on two binary outcomes.

## 3 METHOD

## 3.1 Overview

Two practical and feasible analysis strategies based on Bayesian model averaging that are suitable when the functional forms of the dose-response relationships for efficacy and toxicity are unknown are described here. These strategies differ from approaches previously described; in particular, (a) they do not assume an explicit functional form for either dose-response relationship, (b) they lead to simple, straightforward ways to identify acceptable doses, (c) they allow for substantial response distribution flexibility, and (d) they are focused on response prediction rather than on hypothesis testing or parameter estimation. Both are implemented using the BMA-Mod approach.[24] BMA approaches have been used previously for assessing uncertainty in the estimation of benchmark doses when modeling environmental toxicity impacts. [25, 26]

**3.2 Analysis Strategies**

Strategy 1 uses marginal BMA dose-response models to obtain posterior or posterior predictive distributions of responses to doses within a selected range, expressed graphically as mean curves with simultaneous credible or predictive bounds as illustrated by Figure 1. Acceptable dosage ranges are defined by requiring expected or predicted efficacy responses to exceed a minimum acceptable level, and expected or predicted toxicity responses not to exceed a maximum tolerable level.

Strategy 2 combines efficacy and toxicity outcomes in a clinically meaningful way so that the responses are counts of subjects falling into each of a set of ordered categories of efficacy and tolerability outcomes defined by clinical, regulatory, or commercial considerations.[27, 28] Table 1 illustrates a potential categorization rule.

Table 1. Efficacy/toxicity outcome categories relative to standard of care (SoC)

| Efficacy | Toxicity relative to SoC | | | |
|---|---|---|---|---|
| | Less | Similar | Worse | Intolerable |
| Better than SoC | 1 | 2 | 4 | 5 |
| Similar to SoC | 2 | 3 | 5 | 5 |
| Less than SoC | 4 | 5 | 5 | 5 |
| Unacceptable | 5 | 5 | 5 | 5 |

The observations typically consist of category membership indicators, e.g.,

| Subject | 1 | 2 | … | N |
|---|---|---|---|---|
| Dose | $d_{(1)}$ | $d_{(2)}$ | … | $d_{(N)}$ |
| Outcome Category | $x_1$ | $x_2$ | … | $x_N$ |

where $d_{(i)}$ denotes the dose administered to Subject i, $d_{(i)} \in \vec{d} = (d_1, \ldots, d_M)$. This approach explicitly incorporates subject matter (e.g., clinical and regulatory) expertise and, simultaneously, associations or correlations between the efficacy and toxicity responses.

Acceptable doses are those where the posterior probability of desired outcomes is sufficiently high.

Although neither of the analysis strategies explicitly considers the possible association between the efficacy and toxicity outcomes for reasons mentioned in section 1.2, efficacy and toxicity outcomes are very likely to be associated in reality, e.g., because the doses affect common receptors on different organ systems. Consequently, the data used to illustrate the analyses should reflect the association between efficacy and toxicity responses, if only to evaluate the effect of different levels and patterns of association on the recommended dosage ranges. Copulas provide a convenient way to simulate data with known marginal dose-response relationships and known associations. We emphasize that the copulas are used here only to build association into the simulated efficacy and toxicity responses, and not for the actual analyses.

Remark: Copula models provide a convenient strategy for incorporating association if explicit accommodation for efficacy-toxicity association is needed when the functional forms of the dose-response models are known.[3, 4, 7, 8, 22, 29-31] They may be less practical when the functional forms are unknown because multiple models corresponding to efficacy, toxicity, and their association, each with its own set of parameters, need to be considered. Incorporating copulas in the context of Bayesian Model Averaging (BMA) dose-response modeling[24] would require fitting separate copula models for each pair of potential dose-response models (one for efficacy, a possibly different one for toxicity), with the predictions from these models averaged as in the univariate case. If there are 6 potential dose-response functional forms, then there would be 36 pairs to consider, not including additional analytic and interpretational complexity when the appropriate copula is unknown.

### 3.3 Simulating Data Using Copulas

Copulas are distributions whose domain is the unit square, i.e., functions $\Psi(u, v; \theta)$ indexed by a parameter $\theta$ where $(u, v)$ is a realization of the random vector $(U, V)$ whose components are uniformly distributed in $(0, 1)$. When U and V are independent, the samples from any copula present the same uniform dispersion pattern in the unit square. When U and V are not independent, the copulas present different dispersion patterns, illustrated in Figure 2. Different functions $\Psi$ characterize copula families with different dispersion patterns. The dispersion patterns for the non-Gaussian copulas (Joe, Clayton, Gumbel, Frank) do not display the same tail symmetry as for the Gaussian copula. This means that the dispersion of (say) V values depends on the corresponding U values.

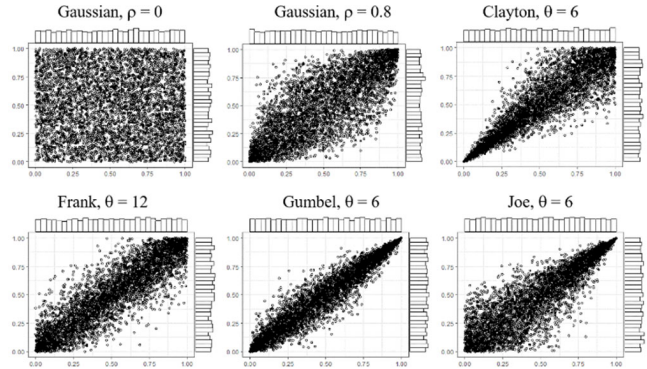

Figure 2.   Bivariate and marginal dispersion patterns for various copulas from samples of 5000

Hofert *et al* [32] and Größer and Okhrin [10] provide comprehensive overviews of copulas with extensive bibliographies. Okhrin *et al* [33] provide software for assessing how well various copulas fit data. The R package '`copula`' [34, 35] provides convenient fuctions for generating values from a variety of copula families, and is used to generate the simulated data for this example.

Simulated 'efficacy' and 'toxicity' outcomes for particular dose-response models were generated using various copula families [10] to generate samples from the unit square (U, V) with a specified correlation or association structure that then were transformed to data points (X, Y) whose distributions were functions of particular dose-response relationships:

$$X = F_E^{-1}(U; \beta_E), \quad Y = F_S^{-1}(V; \beta_S)$$

For example, if X has a normal distribution with standard deviation $\sigma$ and mean function $f_E(\text{dose}; \beta_E)$ of dose, and Y has a Bernoulli distribution with parameter $\pi$ that is another function $f_S(\text{dose}; \beta_S)$ of dose, then the simulated values of X and Y corresponding to U and V are

$$F_E(w; \text{dose}, \beta_E) = \Phi\left(\frac{w - f_E(\text{dose}; \beta_E)}{\sigma}\right) \qquad F_S(z; \text{dose}, \beta_S) = \{1 - \text{expit}(f_S(\text{dose}; \beta_S)\}^{1-z}$$

where $\Phi$ denotes a standard normal cdf and

$$X = \sigma\Phi^{-1}(U) + f_E(\text{dose}; \beta_E) \qquad Y = 1 - \log(V)/\log(1 - f_S(\text{dose}; \beta_S))$$

### 3.4 Differences between Ordered Categories

Various metrics such as utilities or distances can be calculated straightforwardly using the realizations from the posterior expected responses or posterior predicted responses. Similar approaches were used by Thall and Russell [27], Thall *et al* [36], and Zhang and Sargent [28]. Clinical and regulatory considerations outside the scope of the present work deterimine how these utilities might be used to assess efficacy-toxicity tradeoffs.

The expression of differences between distributions of ordered categories is less straightforward than among scalar objects, for which arithmetic differences, ratios, or odds ratios might be used. The MCMC runs produced by the BMA-Mod software provide for each model (and for the weighted model) sets of realizations from the posterior distributions of the category probabilities, Distances between these sets of probabilities for the test doses and the control dose can be calculated for each realization. Symmetric divergence metrics such as the Hellinger or Jensen-Shannon distance measures are useful for evaluating ordered categorical outcomes. [37, 38]

## 4 RESULTS

The calculations were carried out using the current version of the BMA-Mod software. [24] Appendix 1 outlines the computational steps. The code needed to reproduce the analyses reported here is provided in the supplementary information, along with an annotated session log. The simulated copula outcomes were generated separately for each dose value to reflect efficacy-toxicity association levels unaffected by dose values. All models were fit with an initial 4000 iteration warmup and subsequent 5000 iterations with thinning to every fifth iteration. Appendix 2 provides graphical displays showing that the model fits were satisfactory.

### 4.1 Example 1

This example simulates dose-response trials of antihypertensive drugs aimed at finding doses that lower sitting diastolic blood pressure (DBP) from baseline by 5 – 10 mm Hg while controlling the risk of important toxicity. The range of desirable blood pressure lowering is constrained to limit the probability of both inadequate effect and dose-induced hypotension.

The expected reduction in blood pressure from baseline ($y_E$) as a function of dose is assumed to be normally distributed with mean defined by an $E_{max}$ model and $\sigma = 2$, while the occurrence of toxicity ($y_S$) is assumed to have a Bernoulli distribution such that the inverse logit of the rate parameter ($p_S$) follows an exponential distribution,

$$E(y_E; d, \vec{b}_E) = b_{E1} + b_{E2} \times d^{b_{E4}}/(b_{E3} + d^{b_{E4}})$$

$$(\text{logit}(p_S); d, \vec{b}_S) = b_{S1} + b_{S2}e^{b_{S3} \times d}$$

For this example, $\vec{b}_E$ = (4, 15, 0.33, 2.8) and $\vec{b}_S$ = (-10, 5, 0.7). The expected value of $y_E$ lies between 5 and 10 when the dose is between 0.2 and 0.6, and the expected value of $p_S$ is below 0.1 if the dose is less than 0.6. Fifty (50) simulated ($y_E$, $y_S$) values were generated separately for doses = 0, 0.2, 0.4, 0.6, 0.8, and 1 from various copula models: gaussian with $\rho$ = 0 and $\rho$ = 0.8; clayton with $\theta$ = 6; and joe with $\theta$ = 6 using the `copula` function in the 'copula' R package. [34,35]. For Strategy 2, the $p_S$ values were interpreted as expected adverse event severities. The "observed" severity values were realizations from a beta distribution with parameters (1, (1-$p_S$)/$p_S$). Adverse event severity categories were defined as:

"None/Mild" ⇔ severity < 0.05

"Moderate" ⇔ severity ∈ (0.05, 0.25)

"Severe" ⇔ severity > 0.25

**4.1.1 Strategy 1 Results**

BMA model fitting was carried out separately for the efficacy and toxicity simulated outcomes. The resulting weighted average model was used to generate predicted mean response levels, with credible intervals, for a fine grid of doses. These are the quantities displayed in Figures 3 and 4. Figure 3 displays the effect of the copula family and the degree of association on the predictions. The 2.5% lower credible bounds on the difference in change from baseline for each dose from the change on the (zero dose) control do not appear to vary materially with the type of copula and strength of association. There appears to be a modest efffect of copula type and association strength on the upper credible bounds on the toxicity risk. Although the MED (Minimum Effective Dose) and MTD (Maximum Tolerated Dose) vary among the cases considered in Figure 3 because of simulation variability, there is reasonable consistency with respect to the doses that should be carried forward in development.

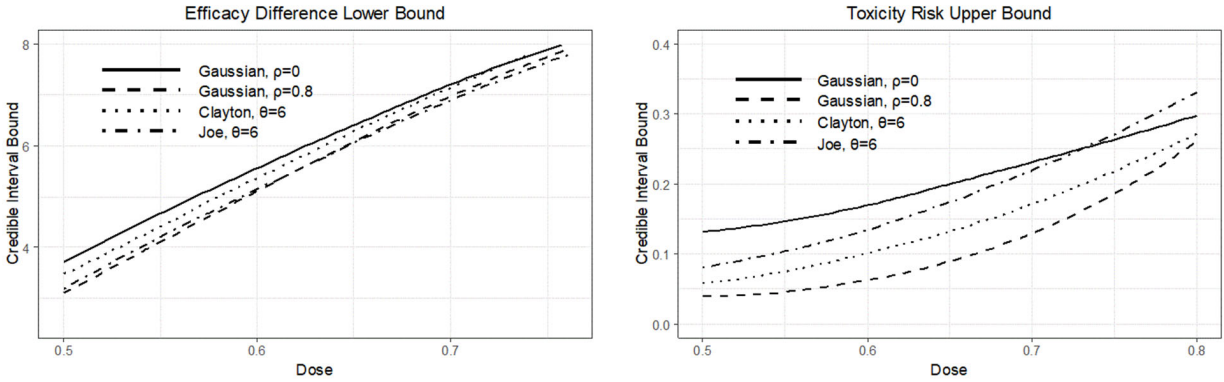

Figure 3.  2.5% lower credible bounds on efficacy difference from control and upper credible bounds on toxicity risk. These bounds are computed separately using the marginal Bayesian weighted average models for efficacy and toxicity.

Figure 4 displays the paired response curves (similar to Figure 1) for this example.

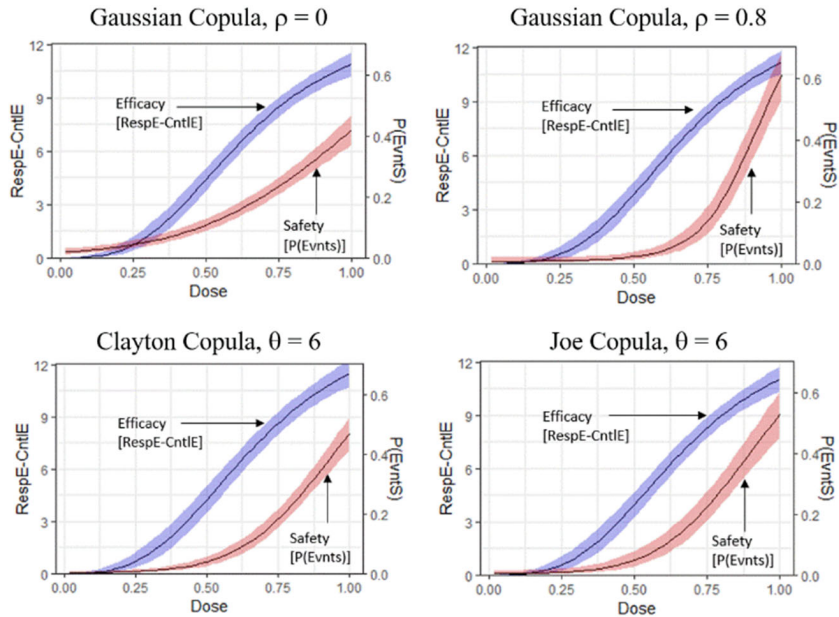

Figure 4.  Paired dose-response curves for response differences from control (zero dose) and toxicity event risks with 95% credible intervals, by degree of association between the efficacy and toxicity responses.

Figure 5 provides an alternative, possibly more intuitive, view of the relationship between the efficacy and toxicity responses as a function of dose: both increase with increasing dose. Figure 5 also suggests which doses lead to acceptable efficacy responses and tolerable toxicity risks. Thus, dose = 0.6 leads to expected responses that are likely to lie between 5 and 10 with toxicity risks less than 0.1, regardless of the copula or degree of association.

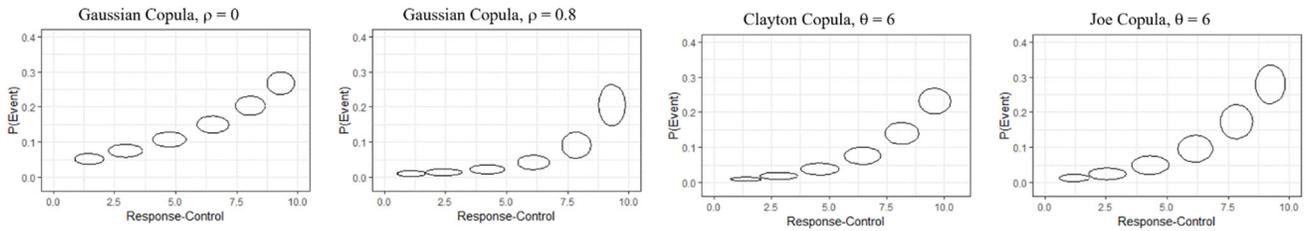

Figure 5. Paired 95% expected dose-response contours for the efficacy differences from control and toxicity event rates as a function of the degree of association of the responses. The contours correspond to doses = 0.3, 0.4, 0.5, 0.6, 0.7, and 0.8

Figures 4 and 5 confirm that the determination of appropriate doses to use going forward that is based on the marginal efficacy and toxicity dose response models does not depend materially on the degree of association or its pattern (copula) between the efficacy and toxicity responses.

### 4.1.2 Strategy 2 Results

Figure 6 displays definitions of clinically meaningful outcome categories applied to the findings for each subject.

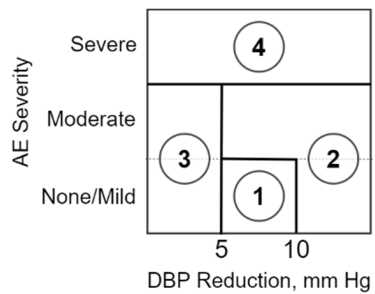

Figure 6. Definitions of outcome categories based on efficacy and toxicity responses

Category 1 ("Best") contains outcomes where the blood pressure reduction is in the target range of 5 to 10 mm Hg and adverse events are absent or at worst mild. Category 2 ("Good") contains outcomes where the blood pressure could be outside of the target range but no less than 5 mm Hg and where the adverse event severity is no worse than moderate. Category 3 ("Minimal") contains outcomes where there is some blood pressure reduction, although less than 5 mm Hg, and where the adverse event severity is no worse than moderate. Category 4 ("Worst") contains outcomes where the adverse event is severe. Analyses were carried out using these categories and also non-disjoint combined categories, "Best", "Best + Good", "Minimal + Worst", "Worst".

The same doses and number of simulated observations per dose were used as for Strategy 1. Predicted probabilities of the various categories were determined using a fine grid of potential doses. Figure 7 displays the dose-response profiles for combinations of categories, including 95% posterior credible intervals for the expected probability corresponding to each response category. The degree of association and, to a lesser effect, the choice of copula, appears to have affected the relationship between the outcome categories and the dose. The effect on the range of 'acceptable' doses does not appear to be material, an observation reinforced by Figure 3.

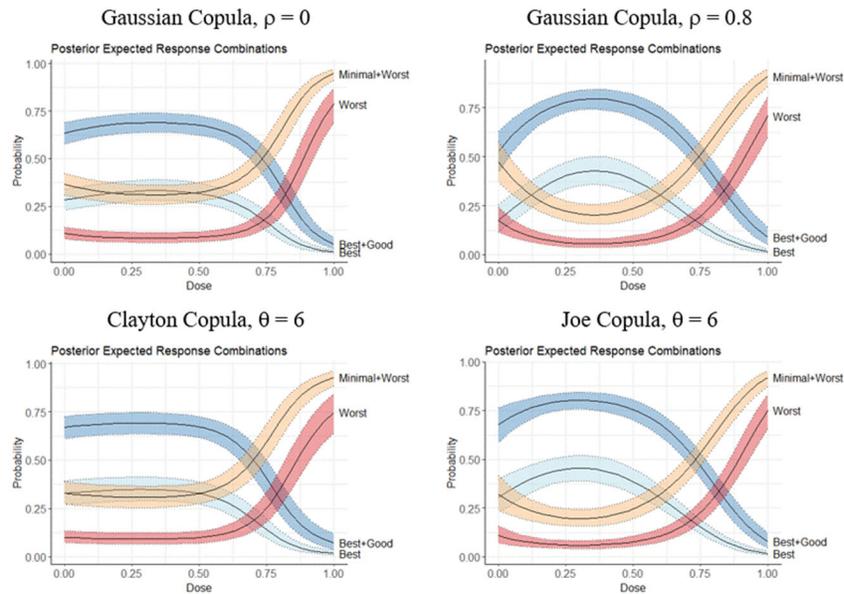

Figure 7. Posterior means and credible intervals for combinations of response categories as a function of the copula generating the simulated response data.

Figure 8 displays the values of 95% lower and upper category-specific credible intervals for doses between 0.2 and 0.8 for the various copula models. Figure 8 includes combined categories, since a "Best" or a "Good" outcome may be desirable if a "Best" outcome is unlikely. For comparison the range suggested from Figure 3 is about 0.58 to 0.64. The ranges in Figure 3 and Figure 8 differ because the objectives regarding the responses differ. Both approaches suggest that a dose between 0.5 and 0.6 would be appropriate regardless of whether the association between the efficacy and toxicity responses (and its value) is incorporated in the analysis or not.

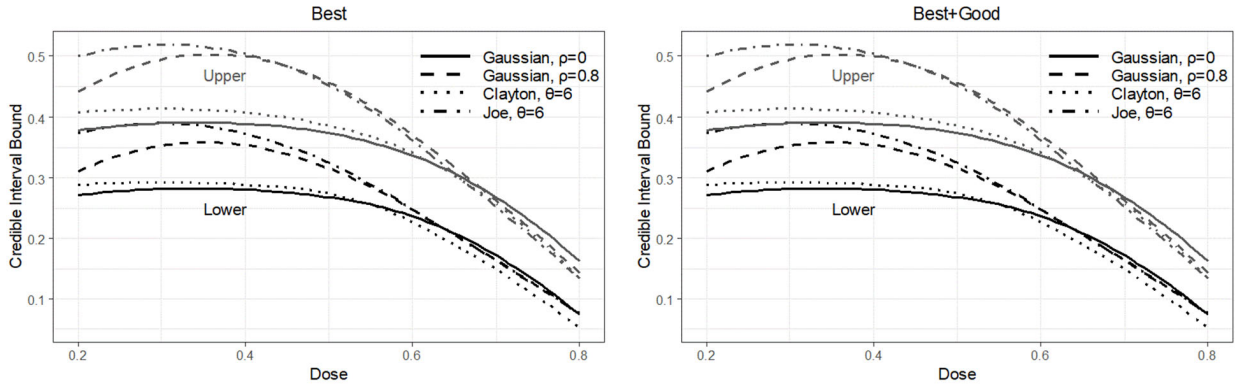

Figure 8.  95% credible intervals for response category probabilities as functions of the copula generating the simulated response data.

Although Figure 8 displays how the distributions among the response categories vary with the dose, how the distributions at the various doses differ from the distribution at the zero dose is not as evident. Figure 9 displays the Hellinger distances [37, 38] between the dose-specific category distributions and the zero dose distribution. The 'optimum' dosage range depends on the strength of the association between the efficacy and toxicity responses and on the correlation/association pattern. Figure 9 suggests that a dose between 0.3 and 0.4 would provide the greatest distributional differences from the zero dose distribution. The lower dose range corresponding to the largest distances reflects primarily differences in efficacy from the zero dose, while the upper dose range reflects primarily differences with respect to toxicity.

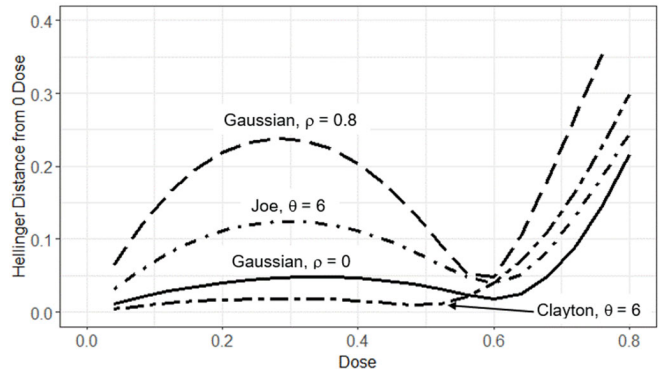

Figure 9.  Hellinger distances between dose-specific category distributions and the zero dose distribution.

### 4.2 Example 2 (Costa and Drury [7])

The simulated efficacy response is assumed to be normally distributed with the expectation following an $E_{max}$ dose-response model,

$$f_E(\text{dose}; \beta_E) = f_E(\text{dose}; \beta_E) = \beta_{E1} + \frac{\beta_{E2} \times \text{dose}}{\beta_{E3} + \text{dose}}$$

where $\beta_{E1} = -150$, $\beta_{E2} = 150$, $\beta_{E3} = 0.5$, and $\sigma = 100$. The simulated toxicity response has a Bernoulli distribution, with the probability p of a toxic outcome following a linear dose-response model via a probit link,

$$p = f_S(\text{dose}; \beta_S) = \Phi(\beta_{S1} + \beta_{S2} \times \text{dose})$$

where $\beta_{S1} = -1.28$ and $\beta_{S2} = 0.26$. The efficacy and toxicity marginal cdfs are

$$F_E(x; \text{dose}, \beta_E) = \Phi\left(\frac{x - f_E(\text{dose}; \beta_E)}{\sigma}\right) \qquad F_S(y; \text{dose}, \beta_S) = p^y(1-p)^{1-y}$$

Fifty efficacy and safety observations were generated at each of the same doses used by Costa and Drury [7] (0, 0.3, 0.7, 1, 4, and 6) using gaussian copulas with $\rho = 0$ and $\rho = 0.8$. A key objective in Costa and Drury [7] is determining doses that will give the highest probability of success, defined in terms of the probability that the expected change in efficacy from the zero dose exceeds 80 units (minimum acceptable efficacy effect) and the expected difference in safety risk is less than 0.3 (maximum tolerable safety risk). Success occurs when the probability for each of these events is at least 0.7.

Expected responses at each of a series of doses were obtained from the weighted models for efficacy and safety[24]. Figure 10 displays the posterior credible intervals for the expected efficacy and toxicity responses as a function of dose, and identifies the doses corresponding to the success requirement. The shaded regions in Figure 10 are the central 40% intervals.

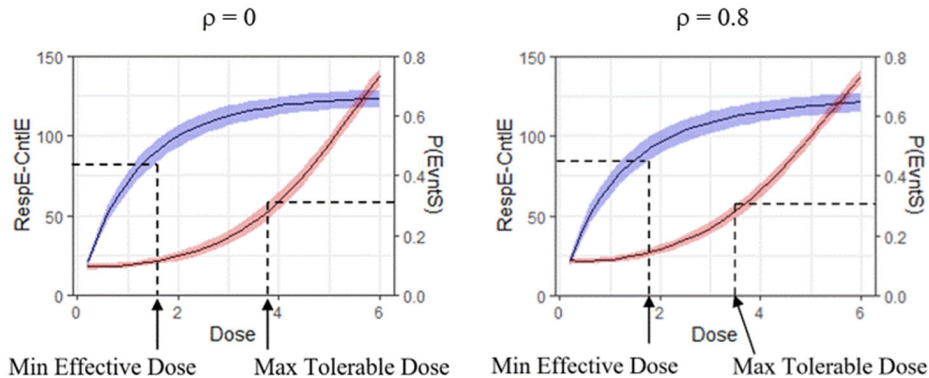

Figure 10. Posterior prediction intervals for efficacy and toxicity responses and differences from control as a function of the administered dose.

The "optimal" dosage range according to Costa and Drury [7] is [2.5, 4]. The corresponding 'optimal' range based on the BMA-Mod calculations is [1.9, 3.6]. The BMA-Mod calculations did not require specifying parametric models for the dose-response relationships.

**4.3 Example 3** (Tao *et al* [4])

This study used an extension of the MCP-Mod approach [21, 39, 40] to determine the Minimum Effective Dose (MED) and the Maximum Safe Dose (MSD). The data consist of changes in glomerular filtration rate as measures of safety and changes in sitting diastolic blood pressure as measures of efficacy. Both outcomes are continuous.

The calculations described here are illustrated using fifty simulated observations of each outcome at doses 0.05, 0.2, 0.4, 0.6, 0.8, and 1. The efficacy outcomes were generated using an $E_{max}$ model for the expected efficacy and an exponential model for the expected safety. The outcomes were generated from a bivariate normal distribution with $\rho = 0$ or $0.8$. The MED and MSD estimates did not appear to depend materially on the degree of correlation. The marginal distributions were

$$F_E(x; \text{dose}, \beta_E) = \Phi\left(\frac{x - f_E(\text{dose}; \beta_E)}{\sigma_E}\right) \qquad F_S(y; \text{dose}, \beta_S) = \Phi\left(\frac{y - f_S(\text{dose}; \beta_E)}{\sigma_S}\right)$$

where

$$f_E(\text{dose}; \beta_E) = \beta_{E1} + \frac{\beta_{E2} \times \text{dose}}{\beta_{E3} + \text{dose}}, \quad \beta_{E1} = 1.4, \beta_{E2} = 14.5, \beta_{E3} = 0.2, \sigma_E = 7$$

$$f_S(\text{dose}; \beta_S) = \beta_{S1} + \beta_{S2} e^{\beta_{S3} \times \text{dose}}, \quad \beta_{S1} = 0.163, \beta_{S2} = 0.037, \beta_{S3} = 5.912, \sigma_S = 8$$

Figure 11 displays the posterior credible intervals for the expected efficacy and toxicity responses as a function of dose, and identifies the doses corresponding to the success requirement. The shaded regions in Figure 11 are the central 95% credible intervals.

The MED is about 0.125 and the MSD is about 0.63, from Figure 11. These values reflect the uncertainty about the true dose-response relationships. If only the mean values (solid black lines) are considered, then the MED and MSD values are 0.06 and 0.75, respectively. It is clear from Figure 11 that the blood pressure does not improve very much past the MSD, while change in glomerular filtration rate increases quite strongly. For comparison, the MED and MSD values reported by Tao *et al* [4] were about 0.06 and 0.8, respectively.

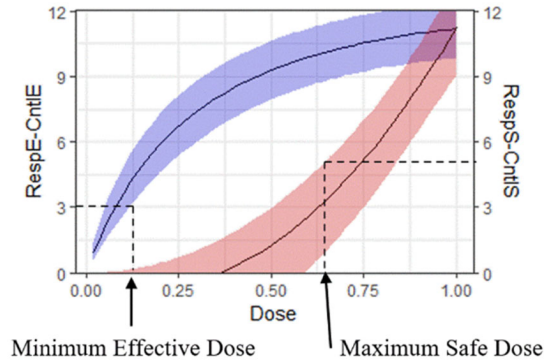

Figure 11. Posterior prediction intervals for differences from control in efficacy (sitting diastolic blood pressure) and toxicity (glomerular filtration rate) responses as a function of the administered dose.

### 4.4 Example 4 (Fukudo et al [41])

Fukudo et al evaluated the effect of three doses of oral lubiprostone for the treatment of constipation with and without irritable bowel syndrome. The primary efficacy outcome was the change from baseline in the weekly average number of spontaneous bowel movements (SBM) at Week 1. A secondary efficacy endpoint was the proportion of patients reporting satisfactory relief after two weeks of treatment. Table 2 summarizes the key findings, including counts of the numbers of patients reporting at least one adverse event during the trial, almost all of which were gastrointestinal symptoms.

Fukudo et al carried out tests of hypotheses comparing the effects of the various dosages against placebo, with adjustments for multiplicity. Only the primary efficacy findings for the two higher dose groups differed significantly from the finding for the placebo group.

Table 2. Efficacy and Safety Findings from Fukudo et al[41]

| Dose, μg | N | Mean Change in SBM | Standard Error | No. with Satisfactory Relief | No. Reporting an AE |
|---|---|---|---|---|---|
| 0 | 42 | 1.5 | 0.4 | 13 | 2 |
| 16 | 41 | 2.3 | 0.4 | 13 | 1 |
| 32 | 43 | 3.5 | 0.5 | 20 | 13 |
| 48 | 44 | 6.8 | 1.1 | 31 | 17 |

Figure 12 presents two graphical displays obtained using the BMA model approach [24] applied to the findings in Table 2. The display in part A of Figure 12 provides estimated dose-response relationships for the efficacy and safety outcomes. The display in part B displays 95% contours for the joint distribution of the safety and efficacy outcomes assuming that they are independent. The findings for the zero dose are not included in Figure 12 because it is apparent from Table 4 that the efficacy and safety findings for the 0 and 16 µg doses were about the same.

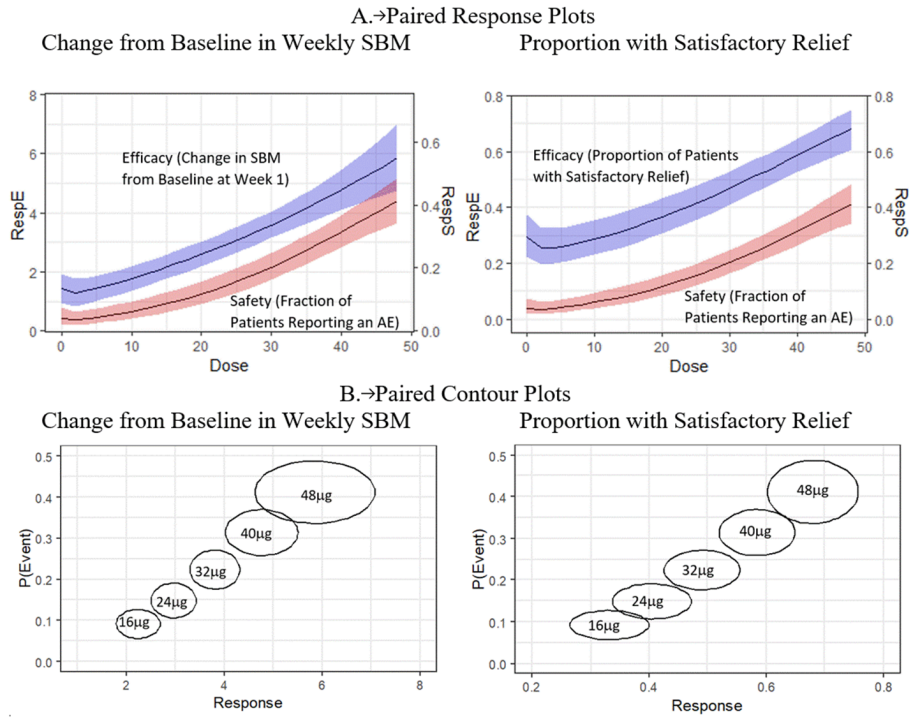

Figure 12. Joint efficacy and safety dose response relationships from Fukudo *et al* [41]

What constitutes an acceptable dose range depends on the definitions of the minimal acceptable clinical response and the maximum tolerable adverse event risk. Figure 12 provides some guidance for determining the range of acceptable doses. The dose range when variability is to be accounted for is narrower than the dose range based only on the expected responses. Doses of 32 µg or more provide better efficacy than a 16 µg dose (and, of course, zero dose). In fact, a 40 µg dose (or perhaps even a 36 µg dose) should provide almost as much clinical benefit as a 48 µg dose, with a substantial reduction in the adverse event risk.

## 5. DISCUSSION

This article describes a novel approach for combining efficacy and toxicity findings from dose-response trials using Bayesian model averaging to identify appropriate doses to carry forward when the true dose-response relationships are unknown. The goal is the prediction of responses or expected responses as a function of dose, rather than the estimation of parameters in a specified mathematical model, or the testing of hypotheses. The method does not require commitment to a particular marginal dose-response model such as an $E_{max}$ model or to a copula model. Uncertainty in the dose-response data used for the analyses translates directly to predictive variability.

The responses can be continuous, counts, binary, or ordinal values, and the distributions of the responses can be normal, student t, Poisson, binomial, Bernoulli, or ordered categorical with various link functions. The models can incorporate stratification and random effects (such as center effects), and baseline covariates. The calculations are carried out using interactive software that requires no coding by the user. Although the approach is illustrated using a single efficacy response and a single safety response, it can be applied without modification to an arbitrary number of efficacy and safety responses, with an appropriate dosage range defined as the intersection of the dosage ranges for each pair of efficacy and safety responses.

Quantities such as the minimum effective dose (MED) the maximum safe or tolerated dose (MSD, MTD) and, most importantly, the range of doses to carry forward, can be read directly from graphical displays and from numerical summaries. In addition, as with the Bayesian averaging method for evaluating dose-response relationships for single responses, a decision as to whether a dose-response relationship exists at all is made separately from the determination of the doses to carry forward based on expected or predicted responses (which do not appear to be obtained easily from methods in the literature).

The method was applied to four examples. The first three contrasted the findings using just the marginal dose-response relationships with the finding using response categories defined by the efficacy and safety responses together. Analyses based on these approaches address different questions and could yield different conclusions. The data used in the calculations were simulated using gaussian and non-gaussian copula models with different association patterns. While the predictions of appropriate dosage ranges were not identical for the marginal and joint model analyses, they were reasonably close and consistent. This suggests that unless the efficacy and safety responses are very strongly associated, the dosage range recommendations are unlikely to depend substantially on the correlation structure. The fourth example illustrated the application of the method to real clinical data, including alternative ways of expressing efficacy. The calculations also suggested that an intermediate dose other than those studied in the trial might be advisable in providing almost the same efficacy and lesser adverse event risk.

Finally, although the proposed method was described and illustrated in the context of determining doses to carry forward in clinical trials, the basic problem it addresses occurs in many other contexts where it is important to determine an input ("dose") to balance a desirable outcome (e.g., "benefit", "yield") with an undesirable one (e.g., "risk", "cost").

## ACKNOWLEGEMENT

**CONFLICTS OF INTEREST**

None

## APPENDIX 1 Outline of BMA-Mod Strategy[24]

Implementation of the BMA-Mod method proceeds in two key steps: (1) Bayesian nonlinear regression analyses using the `brms` R package to produce posterior distributions of model parameters, and (2) calculation of various quantities related to predictive distributions based on the various models. All of the calculations are executed by menu-driven interfaces where the user supplies minimal necessary information.

The nonlinear regression analyses require an input array with one row per observation and columns corresponding to the model variables, of which only two are required:

- X   Response. Can be continuous (Gaussian, Student t), count (Binomial, Poisson), binary (Bernoulli), or ordered category (integer index). Continuous responses can be individual responses or summary statistics with corresponding standard error estimates.
- Dose   A numerical value corresponding to the response.

Additional variables can be specified as required by the model. These include stratum (group) identifiers so that models can be fit separately in each stratum, covariates, or random effects. The random effects and stratum identifiers can apply only to the model intercepts, or to all of the model parameters.

Prior distributions can be specified, e.g., normal, cauchy, student t with 3 df, along with specifications of means and standard deviations. A typical specification is a t(3) distribution with mean zero and standard deviation 5.

The summarization step takes the result of the first step as input and, through a dialogue with the user, produces a variety of calculations and graphics based on the predictive distributions of the responses for the various models and, most importantly, for the predictions from the model obtained as the weighted average of the predictions of the individual models. As with the first step, the user needs to supply only the key information, with all processing done automatically.

## APPENDIX 2     Marginal Model Fit Summaries

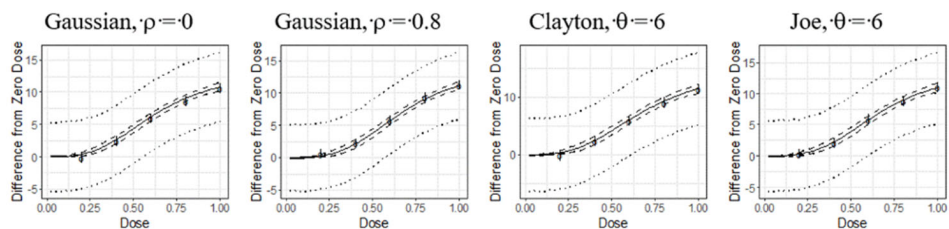

Figure A1 (Example 1). Observed efficacy responses 95% credible intervals, expected responses, credible intervals, and prediction intervals for individual responses.

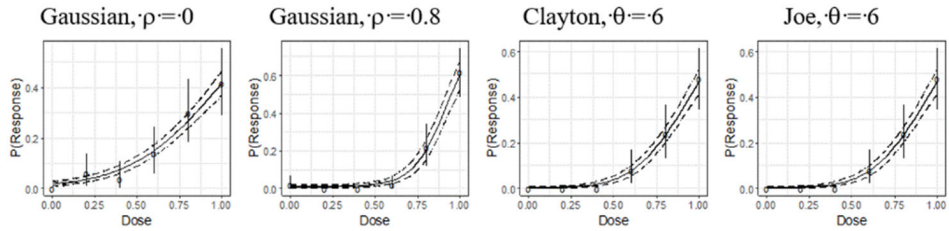

Figure A2 (Example 1).   Observed probabilities of toxicity with 95% credible intervals, expected responses, credible intervals and prediction intervals for responses of individual subjects (which overlap the credible intervals).

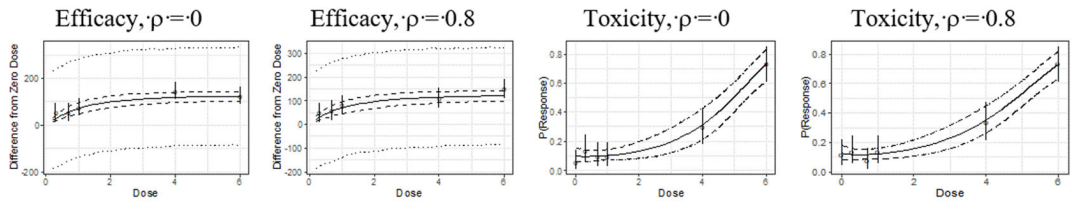

Figure A3 (Example 2)   Observed responses 95% credible intervals, expected responses, credible intervals, and prediction intervals for individual responses.

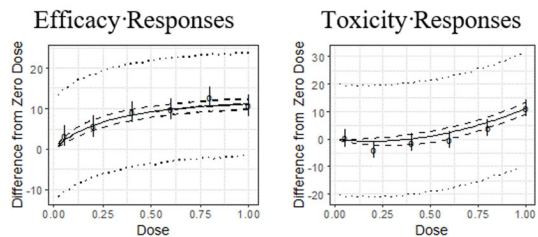

Figure A4 (Example 3)   Observed efficacy and toxicity responses 95% credible intervals, along with predicted responses, credible intervals for the expected responses, and prediction intervals for responses of individual subjects .

## APPENDIX 3   Conditional Dose-Response Models

The unconditional marginal dose-response assessments turn out to be essentially unaffected by the correlation or association between the efficacy and toxicity responses. This is not true when the responses are analyzed together. The conditional dose response relationships may be of interest in some circumstances, especially when a toxic response occurs, and dose reduction may be in order if the adverse event is not intolerable. The following figures display efficacy dose-response relationships for Example 1 when the efficacy responses of the subjects with an adverse event ($y_S = 1$) are considered.

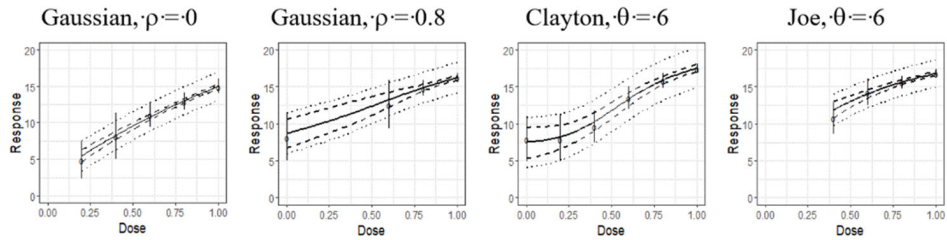

Figure A5. Efficacy dose-response relationships for subjects experiencing an adverse event

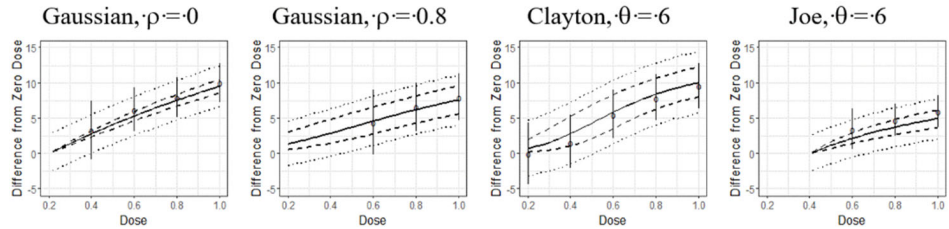

Figure A6. Efficacy dose-response relationships in terms of difference from the zero dose response for subjects experiencing an adverse event

It seems evident from Figures A5 and A6 that the association/correlation between the efficacy and toxicity responses affects the distribution of responses among patients experiencing an adverse event. For example, the expected efficacy responses and differences from the zero dose at dose d=0.6 are

|  | Gaussian, $\rho = 0$ | Gaussian, $\rho = 0.8$ | Clayton, $\theta = 6$ | Joe, $\theta = 6$ |
|---|---|---|---|---|
| Expected response | 10.2 | 12.5 | 13 | 14.5 |
| Diff from 0 dose | 5.5 | 5.0 | 5.1 | 2.6 |